\documentclass[notitlepage,aps,11pt,amsmath,amsfonts,amssymb,nofootinbib]{revtex4-1}
\usepackage{enumerate}
\usepackage{graphicx}

\usepackage{amsthm}

\newtheorem{thm}{Theorem}
\newtheorem{prf}{Proof of Theorem}

\begin{document}

\newcommand{\new}[1]{\ensuremath{\blacktriangleright}#1\ensuremath{\blacktriangleleft}}

\title{Radar orthogonality and radar length in Finsler and metric spacetime geometry}

\author{Christian Pfeifer}
\email{christian.pfeifer@itp.uni-hannover.de}
\affiliation{Institute for Theoretical Physics, Universit\"at Hannover, Appelstrasse 2, 30167 Hannover, Germany}

\begin{abstract}
The radar experiment connects the geometry of spacetime with an observers measurement of spatial length. We investigate the radar experiment on Finsler spacetimes which leads to a general definition of radar orthogonality and radar length. The directions radar orthogonal to an observer form the spatial equal time surface an observer experiences and the radar length is the physical length the observer associates to spatial objects. We demonstrate these concepts on a forth order polynomial Finsler spacetime geometry which may emerge from area metric or pre-metric linear electrodynamics or in quantum gravity phenomenology. In an explicit generalisation of Minkowski spacetime geometry we derive the deviation from the euclidean spatial length measure in an observers rest frame explicitly.
\end{abstract}

\maketitle

%\tableofcontents

%%%%%%%%%%%%%%%%%%%%%%%%%%%%%%%%%%%%%%%%%%%%%%%%%%%%%%%
\section{Introduction}
\begin{figure}[h]
       \centering
        \includegraphics[width=0.14\textwidth]{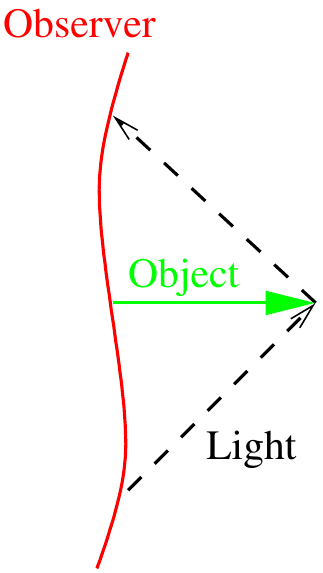}
	\caption{Sketch of a radar experiment: An observer emits light along an object which propagates back to the observer.}
	\label{fig:radarsk}
\end{figure}

%Mathematically this is realised by a length measure $\ell$ which measures the length of the tangent of the light trajectory through space $\dot{\vec{ x}}(t)$, this means $\ell(\dot{\vec{x}}(t))=c$. 
One fundamental axiom of special relativity is that the magnitude of the spatial speed of light is a constant $c$ for every observer. This axiom, which has been experimentally verified to high accuracy, see for example section 2.1.2 in \cite{lrr-2014-4}, leads directly to the fact that observers are able to determine the spatial length of a given comoving object via the following radar experiment. An observer emits a light signal which propagates along the object, the light gets reflected at the edge of the object and propagates back to the observer as sketched in figure~\ref{fig:radarsk}. 
The spatial length $\ell$ the observer associates to the object is now given by half of the time interval $\tau_O$ between the emission and the redetection of the light by the observer multiplied by the constant speed of light $c$
\begin{equation}
\ell(Object)=\frac{\tau_O}{2} c\,.
\end{equation}
Since the propagation of light is encoded in the geometry of spacetime, this experiment intertwines an observers notion of lengths and spatial directions with the spacetime geometry. In general and special relativity the radar experiment is analysed in Minkowski respectively Lorentzian metric geometry where light propagates along the null cone of the metric geometry of spacetime. It yields that the spatial lengths $\ell$ of an object is measured with the spacetime metric and that the orthogonal spatial complement to the observer, i.e. the set of spatial directions which generate the observers equal time surface, is given by orthogonality to the observers tangent with respect to the spacetime metric.

Throughout this article we analyse the radar experiment on generalisations of Minkowski, respectively Lorentzian metric geometry, namely on Finsler spacetimes. Here the geometry of spacetime is not determined by a spacetime metric but by a general length function for vectors. Our goal is to analyse how an observers measurements of length via a radar experiment is intertwined with the spacetime geometry based on such a general length measure. As we will see the analysis of the general Finsler geometric case includes the known results from metric spacetime geometry for a length function $L(Y)=g(Y,Y)$, where $g$ is the Lorentzian spacetime metric. In physics Finslerian spacetime geometries are realised for example in area metric electrodynamics \cite{Punzi:2007di, Schuller:2009hn} and pre-metric electrodynamics \cite{Hehl}, they are considered as extensions of special \cite{Bogoslovski, Cohen:2006ky, Gibbons:2007iu} and general relativity \cite{Rutz, Pfeifer:2011xi} and appear in the studies of quantum gravity phenomenology \cite{Girelli:2006fw, Amelino-Camelia:2014rga}. The study of the non metric electrodynamics demonstrate that it is justified to keep the usual interpretation of the null directions on spacetime also on Finsler spacetimes, namely that these are the directions along which light propagates. The applications as spacetime geometry extending general relativity, or as effective geometry emerging from quantum gravity phenomenology show that it is important to understand how observers measure spatial lengths in Finslerian spacetime geometries. Here we develop a precise formalism which enables us to compare such theories with simple length measurements and to study their deviation from special and general relativity. The introduction of radar lengths and radar orthogonality is only the first step in the analysis of measurable consequence of a Finslerian spacetime geometry. In an ongoing project we consider the other axiom of special relativity, the existence of inertial observers, and study the transformations between different observer on Finsler spacetimes as well as the change in the time dilations and length contractions compared to metric spacetime geometry.

The investigation of the radar experiment for a general Finslerian spacetime geometry leads to the definition of radar orthogonality and radar length without the need of a spacetime metric. Directions radar orthogonal to a timelike observer direction form the spatial complement of the observer, i.e the directions the observer experiences as spatial equal time surface. The radar length obtained via the radar experiment yields the physical spatial length an observer associates to spatial objects. Moreover the introduction of the concepts of radar orthogonality and radar length is also of mathematical interest. The question how to define orthogonality on normed vector spaces without a scalar product has a long history, see \cite{Alonso2012} for a recent overview, and here we introduce a generalised notion of orthogonality on vector spaces equipped not with a norm but with an indefinite length function. 

The structure of the article is as follows. We begin section \ref{sec:1} with a review of the mathematical definitions and properties of Finsler spacetimes in section \ref{sec:fst}, which are the basis for our analysis. Afterwards we introduce the new concepts of radar orthogonality and radar length on Finsler spacetimes in section \ref{sec:def} and analyse their properties in section \ref{sec:prop}. In section \ref{sec:ex} we consider an explicit non metric Finsler spacetime geometry on which we demonstrate that the new concepts are non-trivial. After studying the most general fourth order polynomial Finsler spacetime in section \ref{sec:gen}, we turn to a specific fourth order polynomial Finslerian geometry which can be compared to Minkowski spacetime in section \ref{sec:expl}.  We display the equal time surfaces for different observers and derive how observers measure the spatial lengths of objects in their rest frame. Finally we discuss our result and give an outlook on future projects based on our results in section \ref{sec:disc}.

%%%%%%%%%%%%%%%%%%%%%%%%%%%%%%%%%%%%%%%%%%%%%%%%%%%%%%%
\section{Radar orthogonality and radar length}\label{sec:1}
The analysis of the radar experiment on general Finsler spacetimes leads directly to the definition of radar length and radar orthogonality without the need of a spacetime metric. First we recall the necessary mathematical details of the Finsler spacetime framework to introduce the language we use throughout this article. Afterwards we immediately define the notions of radar orthogonality and radar length and clarify how these describe the radar experiment. In the final part of this section we investigate the mathematical properties of radar orthogonal vectors and their radar length. The results from this chapter demonstrate clearly that it is possible to define spatial orthogonal complements to timelike observers consistently without the need of a metric spacetime geometry. From a mathematical point of view it is surprising that radar orthogonality shares properties like homogeneity with inner product orthogonality without being trivial.

%%%%%%%%%%%%%%%%%%%%%%%%%%%%%%%%%%%%%%%%%%%%%%%%%%%%%%%
\subsection{Finsler spacetimes}\label{sec:fst}
Finsler spacetimes are a very general well defined generalisation of Lorentzian metric manifolds. The difference is that instead of on a metric the geometry is based on a general length function $L$ which associates a number to each vector in the tangent spaces to the spacetime. In the literature there exist various approaches to use Finsler geometry as generalised spacetime geometry, for example in  \cite{FLG, Asanov, Beem}, which all have their advantages and drawbacks. Here we use the definition of Finsler spacetimes from \cite{Pfeifer:2011tk, Pfeifer:2011xi} since it generalises the previous approaches and circumvents the problem of the non existence of the geometry of spacetime along the non-trivial null directions.

To recall the basic properties of Finsler spacetimes we introduce the following notation: a point $Y$ on the tangent bundle $TM$ of a manifold $M$ in some coordinate neighbourhood $V\subset TM$ emerging from a coordinate neighbourhood $U\subset M$ is labeled with coordinates $(x,y)$ and represents the vector $Y=y^a\frac{\partial}{\partial x^a} _{|x}\in T_xM$. These coordinates directly give rise to the coordinate basis of $TTM$ and $T^*TM$ denoted by ${\big\{\partial_a=\frac{\partial}{\partial x^a}, \bar\partial_a=\frac{\partial}{\partial y^a}\big\}}$ respectively $\{dx^a, dy^a\}$. 

\vspace{6pt}\noindent\textbf{Definition 1.}
\textit{A Finsler spacetime $(M,L)$ is a four-dimensional, connected, Hausdorff, paracompact, smooth manifold~$M$ equipped with a continuous function $L:TM\rightarrow\mathbb{R}$ on the tangent bundle which has the following properties:
\begin{enumerate}[(i)]
\item $L$ is smooth on the tangent bundle without the zero section $TM\setminus\{(x,0)\}$;\vspace{-6pt}
\item $L$ is positively homogeneous of real degree $r \ge 2$ with respect to the fibre coordinates of $TM$,
\begin{equation}\label{eqn:hom}
L(x,\lambda y)  = \lambda^r L(x,y) \quad \forall \lambda>0\,;
\end{equation}
\item \vspace{-6pt}$L$ is reversible in the sense 
\begin{equation}\label{eqn:rev} 
|L(x,-y)|=|L(x,y)|\,;
\end{equation}
\item \vspace{-6pt}the Hessian $g^L_{ab}$ of $L$ with respect to the fibre coordinates is non-degenerate on $TM\setminus A$ where~$A$ has measure zero and does not contain the null set $\{(x,y)\in TM\setminus\{(x,0)\}\,|\,L(x,y)=0\}$,
\begin{equation}
g^L_{ab}(x,y) = \frac{1}{2}\bar\partial_a\bar\partial_b L\,;
\end{equation}
\item \vspace{-6pt}the unit timelike condition holds, i.e., for all $x\in M$ the set 
\begin{equation}
\Omega_x=\Big\{y\in T_xM\,\Big|\, |L(x,y)|=1\,,\;g^L_{ab}(x,y)\textrm{ has signature }(\epsilon,-\epsilon,-\epsilon,-\epsilon)\,,\, \epsilon=\frac{|L(x,y)|}{L(x,y)}\Big\}
\end{equation}
contains a non-empty closed connected component $S_x\subset \Omega_x\subset T_xM$.
\end{enumerate}
The Finsler function associated to $L$ is $F(x,y) = |L(x,y)|^{1/r}$ and the Finsler metric $g^F_{ab}=\frac{1}{2}\bar\partial_a \bar\partial_b F^2$.}
The fundamental geometry function $L$ defines the geometry of a Finsler spacetime in a similar way as a Lorentzian metric defines the geometry of a Lorentzian spacetime. All details on the geometry of Finsler spacetimes and their interpretation from a physics viewpoint can be found in the articles \cite{Pfeifer:2011tk, Pfeifer:2011xi}. As a remark observe that for a two homogeneous fundamental geometry function $L(x,y)=g_{ab}(x)y^ay^b$, built from a Lorentzian metric with components $g_{ab}(x)$, Finsler spacetimes are Lorentzian metric spacetimes. 

Most important for the definition of radar orthogonality and radar length are two properties of Finsler spacetimes which follow from requirement $(v)$ of Definition 1. The first one is that $L(x,y)$ defines an indefinite length function for vectors in $T_xM$, i.e. it associates to each vector $Y=y^a\partial_a\in T_xM$ a real number, but is not positive definite
\begin{eqnarray}
L:T_xM&\rightarrow& \mathbb{R}\\
 Y=y^a\partial_a&\mapsto& L(x,y)\,.
\end{eqnarray}
Thus $L$ is not a norm on $T_xM$ and we rather call it an indefinite length function since there exist vectors with positive, negative and null length, similarly as on Lorentzian metric spacetimes. The second important property is the precise definition and existence of timelike vectors. In each tangent space to the manifold it is guaranteed that there exists a cone of vectors $C_x\subset T_xM$, bounded by null vectors, obtained from a rescaling of the shell $S_x$ defined in requirement $(v)$ of Definition 1. The elements of $C_x$ are identified as the timelike and $S_x$ as the unit timelike directions at the point $x$ on the Finsler spacetime $(M,L)$, see \cite{Pfeifer:2011tk} for a proof and more details. These two highlighted properties of Finsler spacetimes enable us to define a notion of orthogonality which defines a spatial complement to each timelike vector in $T_xM$ and the physical spatial length of each element of this complement. We now introduce to the central definitions of this article.

%%%%%%%%%%%%%%%%%%%%%%%%%%%%%%%%%%%%%%%%%%%%%%%%%%%%%%%
\subsection{Definition of radar orthogonality and radar length}\label{sec:def}
In the literature on orthogonality in normed spaces without scalar product exist various generalisation of the usual definition of orthogonality with respect to an inner product, see \cite{Alonso2012} for a modern survey. One possible generalisation of scalar product orthogonality on a normed space, so called isosceles orthogonality, is that an element $u$ of the normed space is orthogonal to another element $v$ if and only if $||u+v||=||u-v||$. The notion of orthogonality we introduce here is a modification of the isosceles orthogonality adapted to the existence of nontrivial null vectors and can be interpreted as the description of the physical radar experiment. Instead of a norm $||\cdot||$ we formulate the orthogonality condition in terms of the fundamental geometry function $L$ which measures the lengths of vectors in the tangent spaces of a Finsler spacetime.

\vspace{6pt}\noindent\textbf{Definition 2.}
\textit{Let $(M,L)$ be a Finsler spacetime and let $U$ be a timelike vector in the tangent space $T_xM$ at $x\in M$. A vector $V\in T_xM$ is called radar orthogonal to $U$ if and only if there exists a scalar $\ell_U(V)$ such that $C=\ell_U(V) U+ V$ and $\tilde C=\ell_U(V) U- V$ are null vectors from the boundary of the cone of timelike vectors $C_x\subset T_xM$
\begin{equation} \label{eq:ortho}
L(C)=L(x, \ell_U(V) U+ V)=0 \text{ and } L(\tilde C)=L(x, \ell_U(V) U- V)=0\,.
\end{equation}
We denote radar orthogonality by the symbol $U\perp_R V$.}

\begin{figure}[h]
       \centering
        \includegraphics[width=0.15\textwidth]{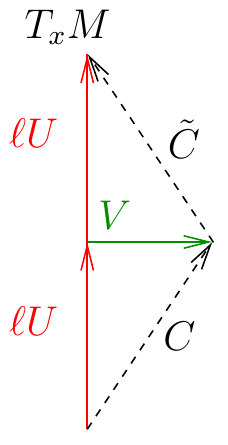}
	\caption{Geometric interpretation of radar orthogonality. For every timelike $U$ there exists a multiple $\ell U$ which can be decomposed into a null vectors $C$ or $\tilde C$ and a spatial vector $V$.}
	\label{fig:1}
\end{figure}

\noindent The geometric idea behind this definition of orthogonality is that the timelike vector $U$ is decomposable into null vectors $C=\ell_U(V) U+ V$ or $\tilde C=\ell_U(V) U- V$ and an orthogonal vector $V$ as sketched in figure \ref{fig:1}. In general it is possible that there exist more than one null cone in the tangent spaces of a Finsler spacetime along which light propagates, see for example the bi-metric example in \cite{Pfeifer:2011tk}, that is why we specify the light cone in the definition of radar orthogonality. On a flat Finsler spacetime, where the tangent space of the spacetime can be identified with the spacetime itself, this situation models a radar experiment in which an observer on a worldline with timelike tangent $U$ emits a light ray with tangent $C$ that gets reflected by a mirror at the end of $V$ and propagates back to the observer along a trajectory with tangent $\tilde C$, as discussed in the introduction and sketched in figure \ref{fig:radarsk}. The function $\ell_U(V)$ is then interpreted as spatial length of $V$ measured by the observer with worldline tangent $U$ in units of time. To convert the length in units of time into the length in conventional length units one multiplies the radar length $\ell_U(V)$ by the constant speed of light $c$. On a curved Finsler spacetime radar orthogonality and radar length still describe the radar experiment infinitesimally, since there exist coordinates on the tangent bundle of a Finsler spacetime around each tangent space $T_xM$ in which curvature effects appear only at second order in the new coordinates \cite{Pfeifer:2014-1}. 

\vspace{6pt}\noindent\textbf{Definition 3.}
\textit{Let $(M,L)$ be a Finsler spacetime and let $U\in T_xM$ be a timelike vector and $V$ be radar orthogonal to $U$. The scalar $\ell_U(V)$ determined from the orthogonality condition is called radar length of $V$ with respect to $U$.}

%\noindent As a remark for the application in physics observe that the radar length has the units  of a time interval. That means to obtain the physical length of an object from its radar lengths one has to divide the radar length by the constant speed of light~$c$.

\vspace{6pt}\noindent\textbf{Definition 4.}
\textit{Let $(M,L)$ be a Finsler spacetime and let $U\in T_xM$ be a timelike vector. We call the set of all radar orthogonal vectors $S(U)$ the spatial complement of $U$ in $T_xM$
\begin{equation}
S(U)=\{V\in T_xM|U\perp_R V\}\,.
\end{equation}}

\noindent The properties of the new notions of radar orthogonality and radar length on Finsler spacetimes can best be studied by a power series expansion of the orthogonality condition (\ref{eq:ortho}) with respect to the components of $V$ and with respect to the components of $U$ in manifold induced coordinates.
\begin{thm}\label{thm:1}
Let $U=U^a\partial_a$ be a timelike vector in $T_xM$ and let $V=V^a\partial_a$ be radar orthogonal to $U$. The orthogonality condition translates into
\begin{eqnarray}
0&=&\sum_{k=0}^\infty \frac{1}{(2k+1)!}V^{a_1}\dots V^{a_{2k+1}}\bar\partial_{a_1}\dots\bar\partial_{a_{2k+1}}L(x,U)(-1)^{2k+1}\ell_U(V)^{-(2k+1)}\label{eq:ortho2}\\
0&=&\sum_{k=0}^\infty \frac{1}{(2k)!}V^{a_1}\dots V^{a_{2k}}\bar\partial_{a_1}\dots\bar\partial_{a_{2k}}L(x,U)\ell_U(V)^{-2k}\label{eq:determinel}\,,
\end{eqnarray}
respectively into
\begin{eqnarray}
0&=&\sum_{k=0}^\infty \frac{1}{(2k+1)!}U^{a_1}\dots U^{a_{2k+1}}\bar\partial_{a_1}\dots\bar\partial_{a_{2k+1}}L(x,V)(-1)^{-(2k+1)}\ell_U(V)^{(2k+1)}\label{eq:ortho3}\\
0&=&\sum_{k=0}^\infty \frac{1}{(2k)!}U^{a_1}\dots U^{a_{2k}}\bar\partial_{a_1}\dots\bar\partial_{a_{2k}}L(x,V)\ell_U(V)^{(2k)}\label{eq:determinel2}\,.
\end{eqnarray}
\end{thm}

\begin{prf}\textnormal{
The proof of Theorem \ref{thm:1} is straightforward. Expand equation (\ref{eq:ortho}) into a power series with respect to the components of $V$, respectively $U$, around $V=0$, respectively $U=0$, use the homogeneity of $L$ and its derivatives with respect to the direction variables to obtain the powers of $\ell$ and realise that the odd and even derivative terms of the expansion have to vanish separately to satisfy the orthogonality condition.} $\square$
\end{prf}
\noindent Observe that for polynomial Finsler spacetimes $L(x,y)=G_{a_1\dots a_r}(x)y^{a_1}\dots y^{a_r}$, which are straightforward generalisations of Lorentzian metric manifolds, the infinite sums in the equations (\ref{eq:ortho2}) to (\ref{eq:determinel2}) terminate after finitely many terms. For $r=2$, a polynomial Finsler spacetime is a Lorentzian metric spacetime and we see directly from the expanded orthogonality condition, equations (\ref{eq:ortho2}) and (\ref{eq:determinel}), that we recover the standard notions of orthogonality and spatial lengths from Lorentzian metric geometry
\begin{eqnarray}
0=V^{a_1}\bar\partial_{a_1}L(x,U)(-1)\ell_U(V)^{-1}&\Rightarrow& G_{a_1a_2}(x)V^{a_1}U^{a_2}=G(U,V)=0\label{eq:metricortho}\\
0=L(x,U)+\frac{1}{2}V^{a_1}V^{a_2}\bar\partial_{a_1}\bar\partial_{a_2}L(x,U)\ell_U(V)^{-2}&\Rightarrow& \ell_U(V)^2=-\frac{G(V,V)}{G(U,U)}\label{eq:metricl}\,.
\end{eqnarray}
Next we investigate further properties of the generalised radar orthogonality and radar length on Finsler spacetimes. 

%%%%%%%%%%%%%%%%%%%%%%%%%%%%%%%%%%%%%%%%%%%%%%%%%%%%%%%
\subsection{Properties of radar orthogonality and radar length}\label{sec:prop}
The radar orthogonality and the radar length introduced in the previous section are defined by two equations, either equations (\ref{eq:ortho2}) and (\ref{eq:determinel}), or equivalently equations (\ref{eq:ortho3}) and (\ref{eq:determinel2}). One of the two equations always determines the spatial length function $\ell$, the other determines the radar orthogonal complement to the timelike reference vector in consideration. Here we want to show that the radar orthogonality and the radar length on Finsler spacetimes share some nice properties with the corresponding notion on Lorentzian metric spacetimes like the scaling behaviour of the radar length $\ell_U(V)$, a kind of non-degeneracy and a homogeneity condition. It is surprising that radar orthogonality has these properties without being trivial since it has been shown that if isosceles orthogonality admits these properties the normed space of consideration must be an inner product space. This is definitely not the case here, as it can be seen from the explicit non-trivial examples we present in the next section. Other properties, like additivity, namely that from $U\perp_R V$ and $U\perp_R W$ follows $U\perp_R (V+W)$, are no longer valid in the general case since the orthogonality condition is no longer linear in its arguments.

\begin{thm}\label{thm:scal}
Let $(M,L)$ be a Finsler spacetime and let $U\in T_xM$ be timelike, $V$ be radar orthogonal to $U$ and $\mu,\lambda\in \mathbb{R}$, then 
\begin{equation}
\ell_{\lambda U}(\mu V)=\frac{\mu}{\lambda}\ell_U(V)\,,
\end{equation}
i.e. $\ell_U(V)$ is homogeneous of degree one in $V$ and homogeneous of degree minus one in $U$.
\end{thm}

\begin{prf}\textnormal{
Consider equation (\ref{eq:determinel}) and write the zeroth order term on the left hand side of the equation to obtain
\begin{equation}
- L(x,U)=\sum_{k=1}^\infty \frac{1}{(2k)!}V^{a_1}\dots V^{a_{2k}}\bar\partial_{a_1}\dots\bar\partial_{a_{2k}}L(x,U)\ell_U(V)^{-2k}\,.
\end{equation}
The left hand side of the equation is $r$-homogeneous with respect to $U$ and zero homogeneous with respect to $V$. To guarantee this scaling behaviour for every term on the right hand side $\ell_U(V)$ must be homogeneous of degree one in $V$ and homogeneous of degree minus one in $U$. The same conclusion can be drawn from equation (\ref{eq:determinel2}). $\square$
}\end{prf}

\begin{thm}
Let $(M,L)$ be a Finsler spacetime and let $U\in T_xM$, then 
\begin{equation}
U\perp_R U \Leftrightarrow L(x,U)=0\,,
\end{equation}
i.e. the only vectors radar orthogonal to themselves are null vectors. In particular this means that no timelike vector can be orthogonal to itself.
\end{thm}

\begin{prf}\textnormal{
If $U\perp_R U$, then by definition $L(x, \ell_U(U) U+ U)=L(x, \ell_U(U) U- U)=0$. By the $r$-homogeneity of $L$ with respect to the direction coordinates $y$ we have
\begin{equation}\label{eq:selfortho}
L(x, \ell_U(U) U+ U)=L(x,U)(\ell_U(U)+1)^r \text{ and }L(x, \ell_U(U) U- U)=L(x,U)(\ell_U(U)-1)^r\,.
\end{equation}
Thus to satisfy the orthogonality condition either $L(x,U)=0$ has to hold or $\ell_U(U)+1=0$ and $\ell_U(U)-1=0$. The later condition can not be fulfilled but be avoided if and only if $\ell_U(U)=0$. The latter inserted in the orthogonality condition again yields $L(x,U)=0$. The other way around, for $L(x,U)=0$ the orthogonality condition is trivially satisfied for $V=U$ by the homogeneity of $L$ as displayed in equation (\ref{eq:selfortho}). $\square$
}\end{prf}
\noindent As a remark observe that the zero vector is radar orthogonal to all timelike vectors $U$,  directly from the defining equation (\ref{eq:ortho}) and from the fact that the zero vector has vanishing radar length $\ell_U(0)=0$. This implies that the spatial complement $S(U)$ always contains the zero vector and is never the empty set. 

\begin{thm}\label{thm:hom}
Let $(M,L)$ be a Finsler spacetime and let $U\in T_xM$ be timelike, $V$ be radar orthogonal to $U$ and $\mu,\lambda\in \mathbb{R}$, then
\begin{equation}
U\perp_R V \Leftrightarrow \lambda U \perp_R \mu V\,,
\end{equation}
i.e. radar orthogonality is homogeneous.
\end{thm}

\begin{prf}\textnormal{
From the scaling behaviour of $\ell_{\lambda U}(\mu V)$, proven in Theorem \ref{thm:scal}, follows the scaling behaviour of the terms in the orthogonality condition
\begin{eqnarray}\label{eq:orthoscal}
L(x, \ell_{\lambda U}(\mu V) \lambda U+ \mu V)&=&L(x,\mu \ell_U( V) U+ \mu V)=L(x,\ell_U(V)U+V)\mu^r \\
L(x, \ell_{\lambda U}(\mu V) \lambda U- \mu V)&=&L(x, \mu \ell_{U}( V)  U- \mu V)=L(x,\ell_U(V)U-V)\mu^r\,.
\end{eqnarray}
Hence 
\begin{eqnarray}
L(x, \ell_{\lambda U}(\mu V) \lambda U+ \mu V)=L(x,\ell_U(V)U+V)\mu^r&=&L(x,\ell_U(V)U-V)\mu^r=L(x, \ell_{\lambda U}(\mu V) \lambda U- \mu V)\nonumber \\
\Leftrightarrow L(x,\ell_U(V)U+V)&=&L(x,\ell_U(V)U-V)\,.
\end{eqnarray}
Thus the orthogonality conditions $L(x, \ell_{\lambda U}(\mu V) \lambda U+ \mu V)=L(x, \ell_{\lambda U}(\mu V) \lambda U- \mu V)=0$ and $L(x,\ell_U(V)U+V)=L(x,\ell_U(V)U-V)=0$ are equivalent. $\square$
}\end{prf}

Having discussed the basic properties of radar orthogonality and radar lengths we now turn to an explicit non-trivial example Finsler spacetime geometry to demonstrate that these concepts are indeed useful and non-trivial.

%%%%%%%%%%%%%%%%%%%%%%%%%%%%%%%%%%%%%%%%%%%%%%%%%%%%%%%
\section{The forth order polynomial Finsler spacetimes}\label{sec:ex}
To demonstrate that the abstract definitions of radar orthogonality and radar length are non-trivial we derive these notions on an explicit non-metric example geometry, which is based on a fourth order polynomial fundamental geometry function $L$. We will compare the results with the standard notions of orthogonality and spatial length known from metric geometry, where the fundamental geometry function is a second order polynomial. First we discuss the general fourth order polynomial case before we turn our attention to a specific flat fourth order polynomial geometry that can be compared to Minkowski spacetime geometry. 

%%%%%%%%%%%%%%%%%%%%%%%%%%%%%%%%%%%%%%%%%%%%%%%%%%%%%%%
\subsection{The general case}\label{sec:gen}
A suitable example of a Finsler spacetime $(M,L)$ beyond metric geometry is the forth order polynomial geometry derived from a fundamental geometry function of the form
\begin{equation}
L(x,y)=G_{abcd}(x)y^{a}y^{b}y^{c}y^{d}=G(y,y,y,y)\,.
\end{equation}
The infinite sums from Theorem \ref{thm:1} which represent the orthogonality condition reduce to the following equations
\begin{eqnarray}
0&=&-  V^a\bar\partial_a L(x,U)\ \ell_U(V)^{-1}-\frac{1}{3!}V^aV^bV^c \bar\partial_a\bar\partial_b\bar\partial_c L(x,U)\ \ell_U(V)^{-3}\nonumber\\
&=&-4( G(V,U,U,U) \ell_U(V)^{-1}+ G(V,V,V,U) \ell_U(V)^{-3})\label{eq:4orderortho}\\
0&=& L(x,U)+\frac{1}{2!}V^aV^b \bar\partial_a\bar\partial_b L(x,U)\ \ell_U(V)^{-2}+\frac{1}{4!}V^aV^bV^cV^d \bar\partial_a\bar\partial_b\bar\partial_c \bar\partial_dL(x,U)\ \ell_U(V)^{-4}\nonumber\\
&=&G(U,U,U,U)+6 G(V,V,U,U)\ell_U(V)^{-2}+G(V,V,V,V)\ell_U(V)^{-4}\label{eq:4orderl}\,.
\end{eqnarray}
Solving equation(\ref{eq:4orderl}) for $\ell_U(V)^2$ yields
\begin{equation}\label{eq:ell}
\ell_U(V)^2=-3\frac{G(V,V,U,U)}{G(U,U,U,U)}\pm\sqrt{9\frac{G(V,V,U,U)^2}{G(U,U,U,U)^2}-\frac{G(V,V,V,V)}{G(U,U,U,U)}}
\end{equation}
and can be used to eliminate $\ell_U(V)^2$ in (\ref{eq:4orderortho}) to obtain a necessary condition on vectors $V$ to be radar orthogonality to $U$
 \begin{eqnarray}\label{eq:nec}
%\ell_U(V)^2&=&-\frac{G(V,V,V,U)}{G(V,U,U,U)}=-3\frac{G(V,V,U,U)}{G(U,U,U,U)}-\sqrt{9\frac{G(V,V,U,U)^2}{G(U,U,U,U)^2}-\frac{G(V,V,V,V)}{G(U,U,U,U)}}\label{eq:4orderl2}\\
0&=&G(U,U,U,U)G(V,V,V,U)^2+G(V,V,V,V)G(V,U,U,U)^2\nonumber\\
&-&6G(V,V,U,U)G(V,V,V,U)G(V,U,U,U)\label{eq:4orderortho2}\,.
\end{eqnarray}
These conditions replace the well known spatial length and orthogonality condition from metric geometry displayed in equation (\ref{eq:metricortho}) and (\ref{eq:metricl}). Equation (\ref{eq:ell}) shows explicitly the scaling behaviour of $\ell_U(V)$ which was derived in Theorem \ref{thm:scal}. Observe that if $V$ solves the necessary condition (\ref{eq:nec}) and $G(V,U,U,U)$ as well as $G(V,V,V,U)$ are non vanishing the only consistent solution for $\ell_U(V)^2$ in equation (\ref{eq:ell}) is the one with the plus sign. The solution with the minus sign is not consistent with equation (\ref{eq:4orderortho}). Since we seek for one solution for all possible $V$ this singles out the physical length an observer assigns to spatial directions as the solution with the plus sign. 

To be even less abstract we now study the concept of radar orthogonality and radar length on a Finsler spacetime geometry which can be compared to Minkowski spacetime.

%%%%%%%%%%%%%%%%%%%%%%%%%%%%%%%%%%%%%%%%%%%%%%%%%%%%%%%
\subsection{An explicit example}\label{sec:expl}
\begin{figure}[h]
       \centering
        \includegraphics[width=0.3\textwidth]{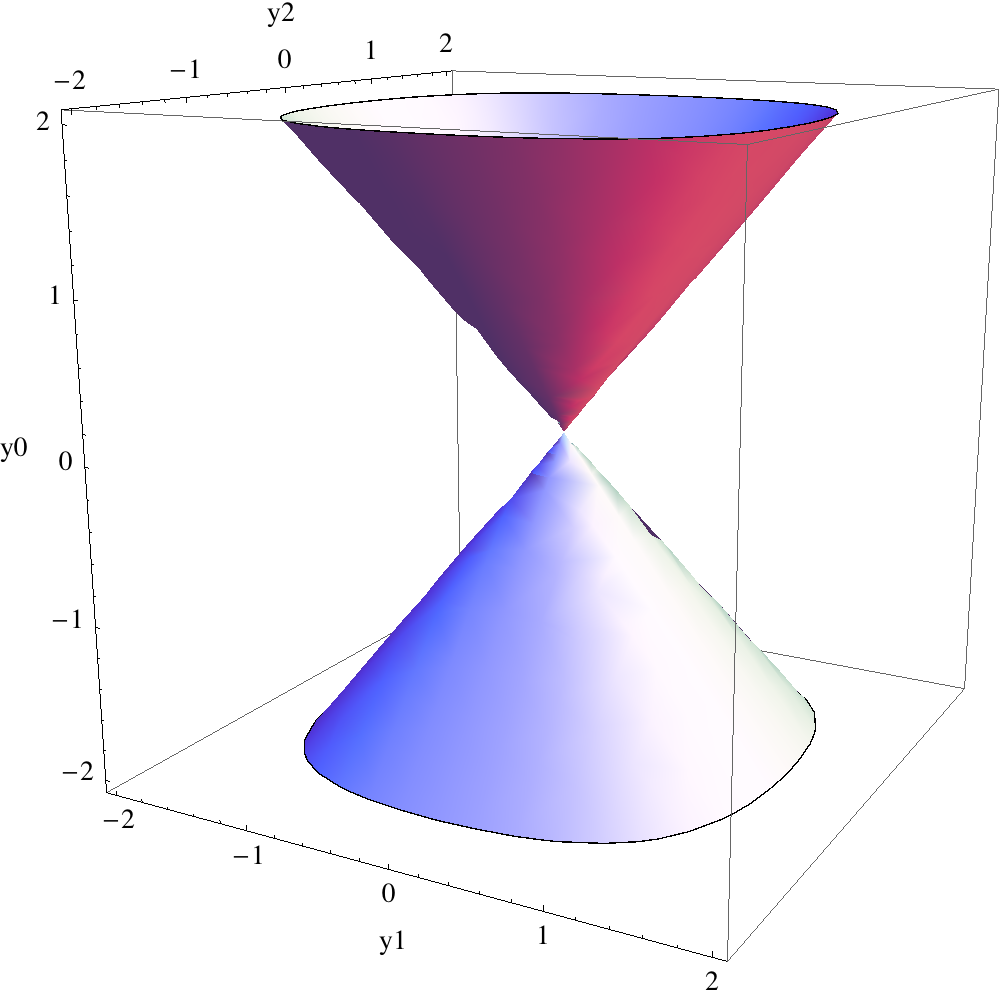}\qquad
        \includegraphics[width=0.3\textwidth]{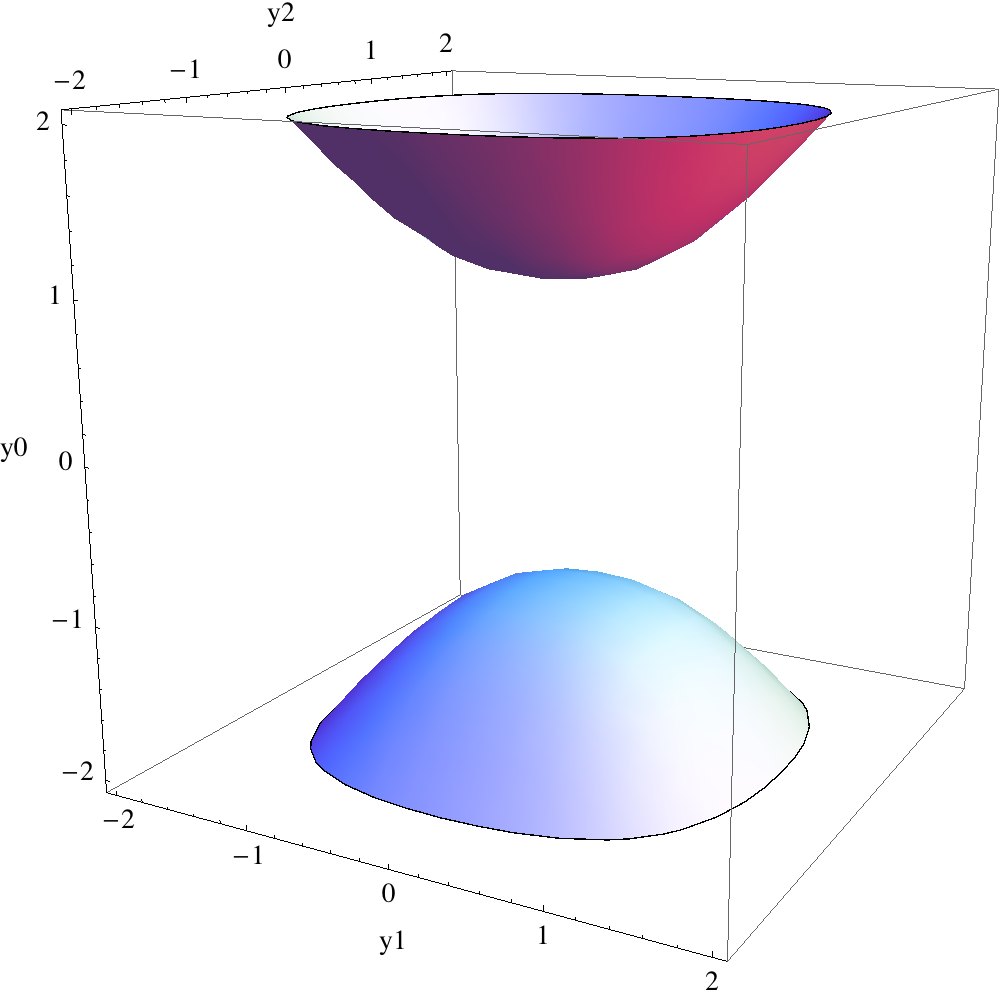}\qquad
        \includegraphics[width=0.3\textwidth]{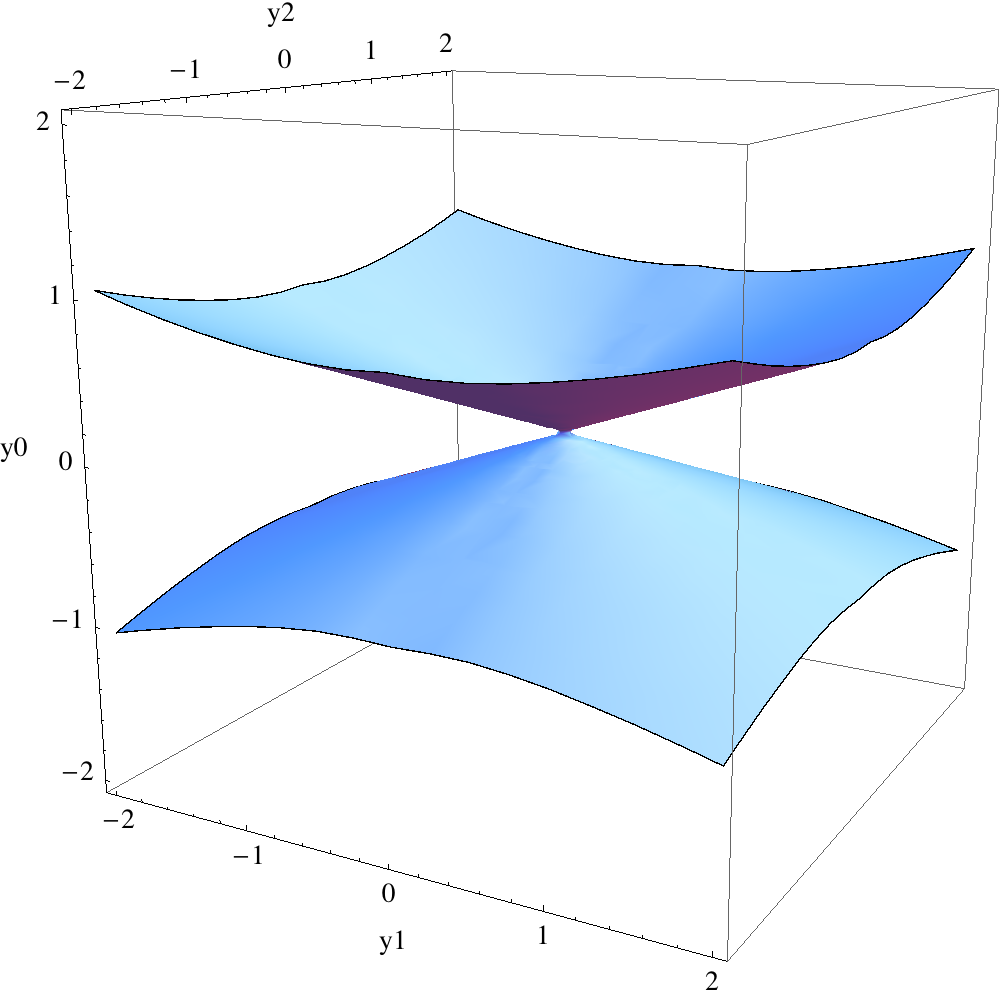}
	\caption{Null cone (left), future and past pointing unit timelike vectors (middle) and the set $A$ along which $g^L$ degenerates (right) of the Finsler spacetime geometry defined by equation (\ref{eq:Lex}).}
	\label{fig:feat}
\end{figure}

Here we study the concepts of radar orthogonality and radar length on a fourth order polynomial Finsler spacetime that is comparable to Minkowski spacetime. To be able to display our results graphically we restrict ourself to a three dimensional Finsler spacetime, all figures have been obtained with Mathematica. Let $M=\mathbb{R}^3$ and introduce coordinates $(x^0, x^1, x^2, y^0, y^1, y^2)$ on $TM$ such that the fundamental geometry function we are interested in takes the following form
\begin{equation}\label{eq:Lex}
L(x,y)=-(y^0)^4+(y^1)^4+(y^2)^4+(y^0)^2((y^1)^2+(y^2)^2)\,.
\end{equation}
This fundamental geometry function indeed defines a fourth order polynomial Finsler spacetime: it is smooth, homogeneous of degree $4$ and reversible, it defines a flat Finsler spacetime since it does not depend on any coordinates of the base manifold and the components of the fundamental tensor $G_{abcd}$ are given by
\begin{equation}
G_{abcd}=\frac{1}{24}\bar\partial_a\bar\partial_b\bar\partial_c\bar\partial_d L\,.
\end{equation}
The important features of the spacetime geometry are depicted in figure \ref{fig:feat}: the null structure in each tangent space is a quartic light cone, the shell of unit timelike vectors lies inside this null cone and the set $A$ along which $g^L$ degenerates lies outside the set of timelike vectors as required
Observe that the degeneracy set lies outside of the set of timelike vectors due to the cross term $(y^0)^2((y^1)^2+(y^2)^2)$ in (\ref{eq:Lex}). This is why we do not consider the maybe simplest fourth order polynomial geometry $-(y^0)^4+(y^1)^4+(y^2)^4$ since in this case the degeneracy of $g^L$ lies inside the cone of timelike vectors.  

\begin{figure}[h!]
       \centering
        \includegraphics[width=0.75\textwidth]{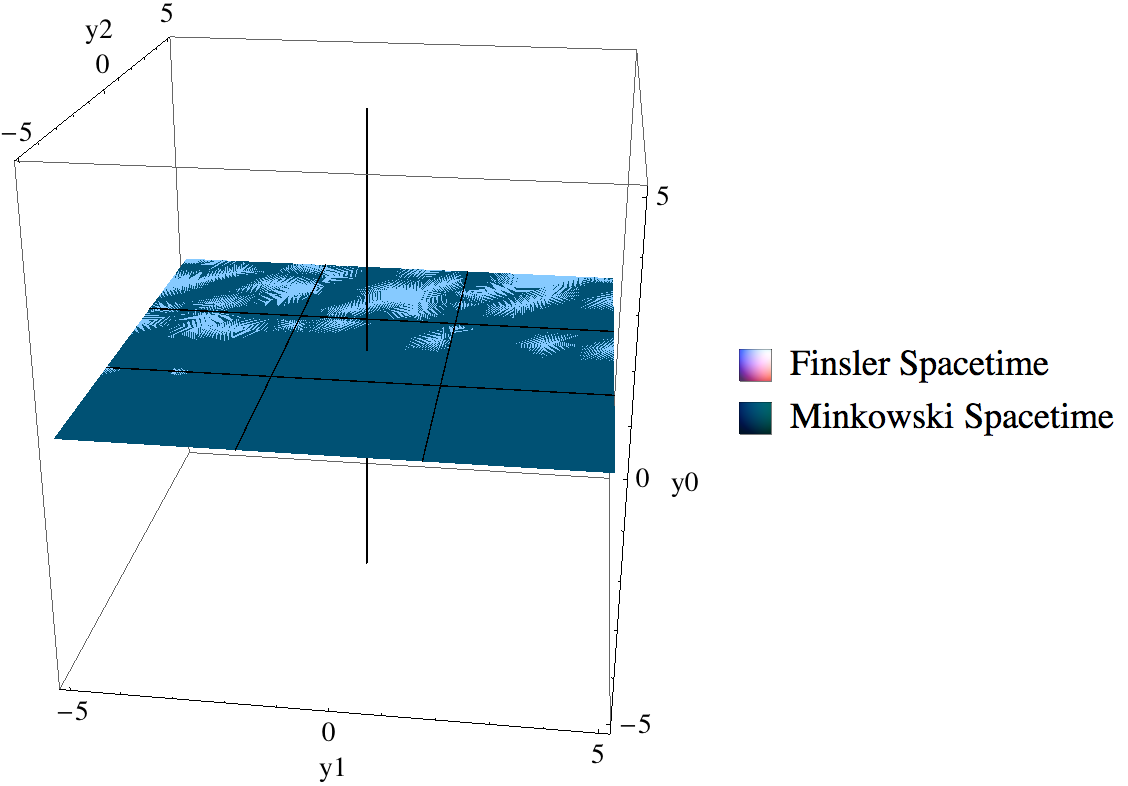}
	\caption{Orthogonal complement to: $U_1$ and $U_{1M}$ including the trajectory $\lambda U_1$.}
	\label{fig:equal1}
\end{figure}

\begin{figure}[h!]
       \centering
        \includegraphics[width=0.75\textwidth]{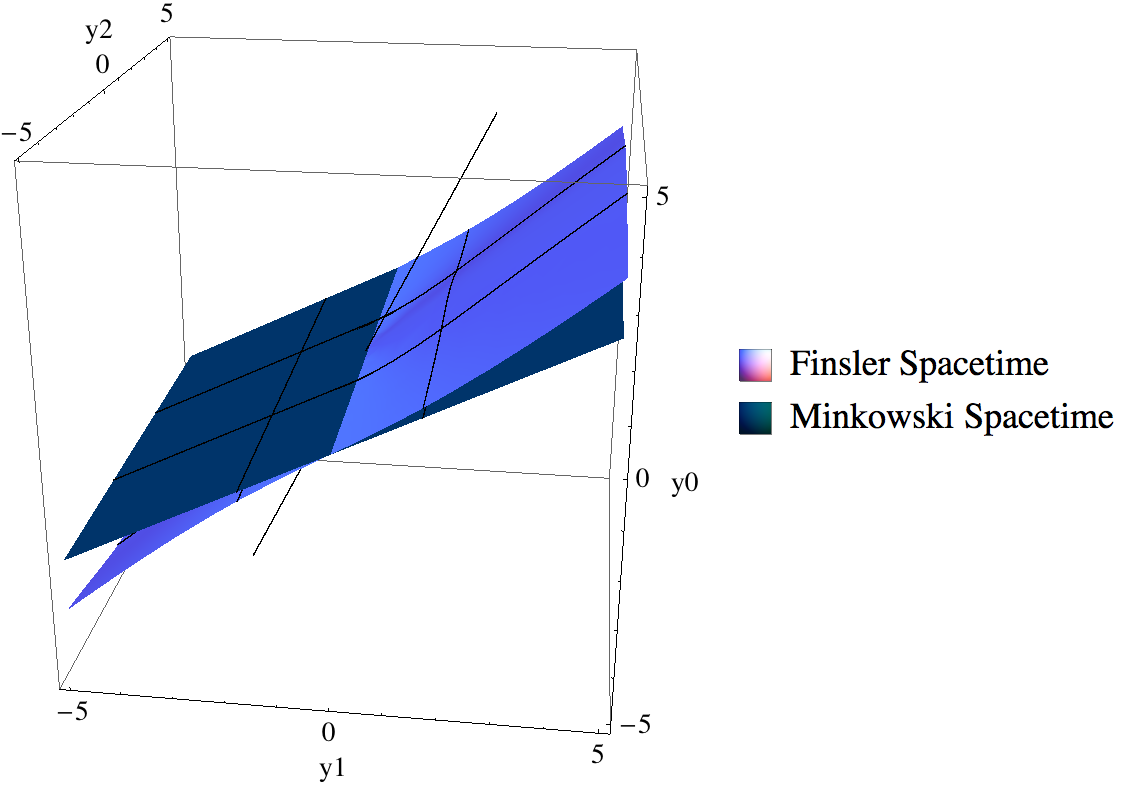}
	\caption{Orthogonal complement to $U_2$ and $U_{2M}$ including the trajectory $\lambda U_2$.}
	\label{fig:equal2}
\end{figure}

\begin{figure}[h!]
       \centering
        \includegraphics[width=0.75\textwidth]{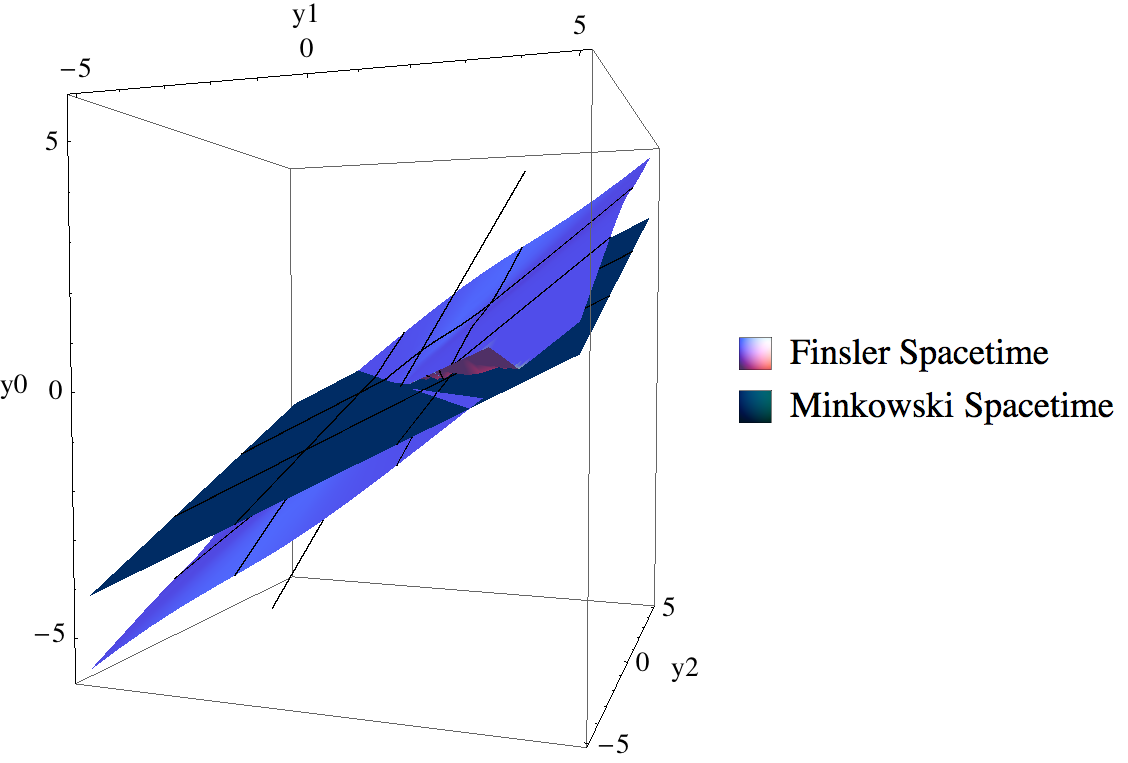}
	\caption{Orthogonal complement to $U_3$ and $U_{3M}$ including the trajectory $\lambda U_3$.}
	\label{fig:equal3}
\end{figure}

We will now display the orthogonal complement of the spatial vector for observers with unit timelike tangents
\begin{equation}
U_1=(1,0,0),\quad U_2=\Big(\frac{16}{11}\Big)^{\frac{1}{4}}\Big(1,\frac{1}{2},0\Big),\quad U_3=\Big(\frac{1296}{731}\Big)^{\frac{1}{4}}\Big(1,\frac{1}{2},\frac{1}{3}\Big)
\end{equation}
and compare them with the orthogonal complements and their lengths of the corresponding observers in a Minkowski spacetime. The corresponding observers in Minkoswki spacetime are considered with the correct normalisation with respect to the Minkowski length element $L_M=-(y^0)^2+(y^0)^1+(y^0)^2$
\begin{equation}
U_{M1}=(1,0,0),\quad U_{M2}=\Big(\frac{4}{3}\Big)^{\frac{1}{2}}\Big(1,\frac{1}{2},0\Big),\quad U_{M3}=\Big(\frac{36}{23}\Big)^{\frac{1}{2}}\Big(1,\frac{1}{2},\frac{1}{3}\Big)\,.
\end{equation}
To obtain the radar orthogonal complement to the unit timelike tangents $U_i$ we first solve the necessary condition all radar orthogonal vectors $V$ have to satisfy, equation (\ref{eq:nec}). Among the solutions there are several which are imaginary and so can be discarded. The real solutions then are checked to solve condition (\ref{eq:4orderortho}) together with the solution 
\begin{equation}\label{eq:radarl2}
\ell_U(V)^2=-3\frac{G(V,V,U,U)}{G(U,U,U,U)}+\sqrt{9\frac{G(V,V,U,U)^2}{G(U,U,U,U)^2}-\frac{G(V,V,V,V)}{G(U,U,U,U)}}
\end{equation}
for $\ell$ from equation (\ref{eq:ell}). Moreover we ensured that the real solutions indeed recombine to null vectors as required in the original radar orthogonality condition (\ref{eq:ortho}). Since the necessary condition of the vectors $V$, equation (\ref{eq:nec}), is a sixth order polynomial in $V$ the calculations could not be performed analytically but only numerically with Mathematica. The spatial orthogonal equal time surfaces of the observers with tangent $U_i$ and $U_{Mi}$ we obtained are displayed in figures \ref{fig:equal1} to \ref{fig:equal3}.
Qualitatively the same features are visible in the metric and the Finsler geometry case. From the standard notion of orthogonality for the observer at rest in figure \ref{fig:equal1}, the orthogonal complement is a hypersurfaces  that tilts towards the observers tangent for observers moving relative to the observer at rest figure \ref{fig:equal2} and figure \ref{fig:equal3}. The main difference between the Finsler case and the metric case is the magnitude of the tilt and that in the metric geometry case the orthogonal complement is a hyperplane, i.e. a sub vector space, while in the Finsler case the orthogonal complement of the moving observers becomes a curved hypersurface, i.e. a submanifold. 

\begin{figure}[h!]
       \centering
        \includegraphics[width=0.8\textwidth]{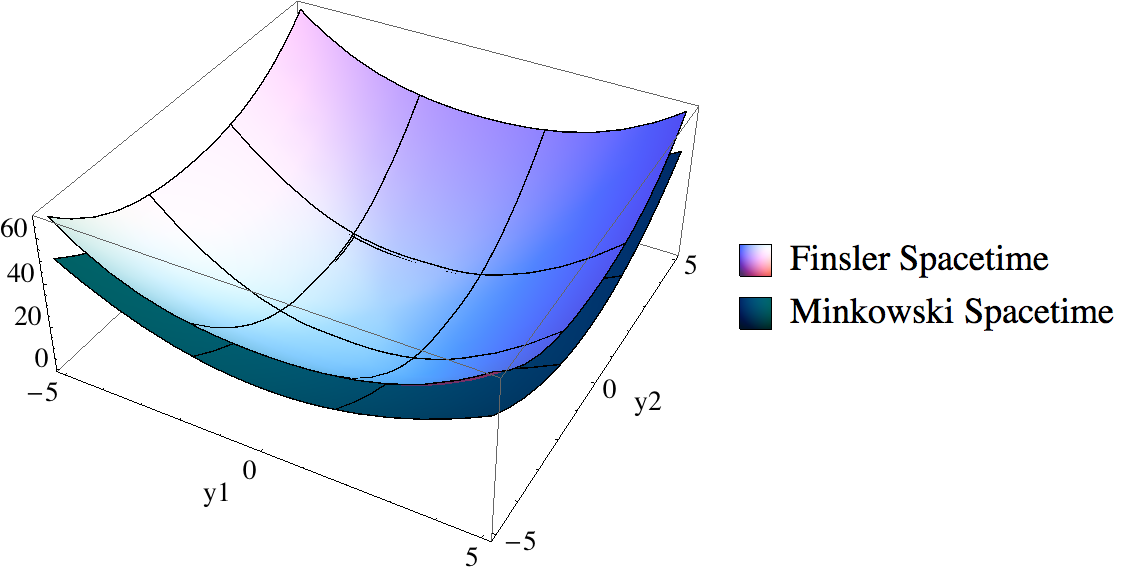}
	\caption{Radar lengths of objects in an observers rest frame on the Finsler spacetime and on Minkowski spacetime.}
	\label{fig:length}
\end{figure}

Finally we compare the radar length of spatial objects in observers rest frames in the Finsler spacetime in consideration , given by equation (\ref{eq:radarl2}), and in Minkowski spacetime, given by equation (\ref{eq:metricl}) with $G$ being the Minkowski metric. Evaluating equation (\ref{eq:radarl2}) on the vectors radar orthogonal to $U_1$, i.e. $V=(0,v^1,v^2)$, and evaluating (\ref{eq:metricl}) on the vectors radar orthogonal to $U_{1M}$, again $V=(0,v^1,v^2)$, yields by direct calculation
\begin{eqnarray}
\ell_{U_1}(V)^2&=&\frac{1}{2}\Big((v^1)^2+(v^2)^2+\sqrt{5(v^1)^4+2 (v^1)^2(v^2)^2+5(v^2)^4}\Big)\label{eq:devieucl}\\
\ell_{U_{1M}}(V)^2&=&(v^1)^2+(v^2)^2\,.
\end{eqnarray}
Thus an observer at rest describing a spatial object with the spatial vector $V_O=(0,1,0)$ associates the spatial length $\ell_{U_1}(V_O)^2=1/2(1+\sqrt{5})$ to the object, while on Minkowski spacetime the observer obtains the length $\ell_{U_{1M}}(V_O)^2=1$. In figure \ref{fig:length} the two different lengths functions are displayed as function of the components of different spatial vectors $V=(0,v^1,v^2)$.
As for the radar orthogonality we observer qualitatively the same behaviour of the length function but quantitative different values for the length of the spatial vector.  The study of the length of different spatial vectors measured in an observers rest frame on the Finsler spacetime (\ref{eq:Lex}) and Minkowski spacetime completes the demonstration that the concepts of radar orthogonality and radar length are non-trivial and enable us to compare predictions from theories with a Finslerian length measure and special, respectively general relativity. 
 
%%%%%%%%%%%%%%%%%%%%%%%%%%%%%%%%%%%%%%%%%%%%%%%%%%%%%%%
\section{Discussion}\label{sec:disc}
In any mathematical theory of the physical world it is necessary to describe how physical observers measure spatial lengths. One practical way to realise this measurement is to consider a radar experiment, as we have done it throughout this article. To be able to describe this experiment for a huge variety of situations like area metric electrodynamics, pre-metric electrodynamics, extensions of special and general relativity as well as results from quantum gravity phenomenology we investigated the experiment in a Finslerian spacetime geometry setting. The main technical result of this article is the introduction of the concepts of radar orthogonality in Definition 2 and radar length in Definition 3 and the fact that this generalised notion of orthogonality is homogeneous in this general Finsler geometric setting, see Theorem \ref{thm:hom}. The central result for physics is that if a spacetime geometry is described by a non-metric Finslerian spacetime geometry, then the spatial length an observer associates to objects follows a different behaviour than on metric spacetime. We illustrated this fact explicitly in the examples discussed in section \ref{sec:ex} where we considered fourth order polynomial Finsler spacetimes. To get an insight how the spatial length measure could change due to the non metricity of the spacetime geometry we finally discussed a specific flat Finsler spacetime geometry on which we derived the deviation from the euclidean spatial length in equation (\ref{eq:devieucl}). 

This work demonstrates that  physical theories based on or leading to a Finsler spacetime geometry share at least one axiom with special relativity. In all of these theories we can implement the axiom of a constant speed of light and, as demonstrated here, analyse the consequences on the measurement of spatial lengths for observers. In an ongoing work we investigate the other axioms of special relativity in the context of a Finslerian spacetime geometry. We will study the existence of inertial observers, the transformations between the observers which will be modifications of the Lorentz transformations and we will calculate the resulting modifications in the time dilation and length contraction due to the non-metric spacetime geometry.

A concrete important future application of our findings is to calculate the effect of first order quantum electrodynamics corrections on the classical metric length measure. As discussed in \cite{Schuller:2009hn} based on the results of \cite{PhysRevD.22.343} the one loop quantum electrodynamics correction to the photon action of Maxwell electrodynamics on a Ricci flat spacetime, i.e. on a Lorentzian metric spacetime which is a solution of the Einstein vacuum equations, yields an area metric electrodynamics with area metric $G^{abcd}=g^{a[c}g^{b]d}+\lambda W^{abcd}$, where $W^{abcd}$ is the Weyl tensor of the spacetime and $\lambda =\alpha/(90\pi m_e^2)$. The analysis of the radar experiment in the Finsler spacetime geometry which describes the propagation of light for this area metric geometry will yield a correction to the metric spatial length measure which comes from first order quantum electrodynamic effects. 

%%%%%%%%%%%%%%%%%%%%%%%%%%%%%%%%%%%%%%%%%%%%%%%%%%%%%%%
\acknowledgments I thank Volker Perlick and Manuel Hohmann for inspiring discussions and remarks. Furthermore I gratefully acknowledge financial support from the Institute for Theoretical Physics of the Leibniz University of Hannover and the hospitality at the ZARM Institute of the University of Bremen.

%%%%%%%%%%%%%%%%%%%%%%%%%%%%%%%%%%%%%%%%%%%%%%%%%%%%%%%
\bibliographystyle{utphys}
\bibliography{test}

\end{document}